\documentclass[preprint,showpacs,preprintnumbers,amsmath,amssymb]{revtex4}

\usepackage{graphicx}% Include figure files
\usepackage{dcolumn}% Align table columns on decimal point
\usepackage{bm}% bold math

\begin{document}

%\preprint{}

\title{Maximally Symmetric Minimal Unification Model SO(32) \\ with Three Families in Ten Dimensional Space-time}
% Force line breaks with \\
\author{Yue-Liang Wu }
 %\email{ylwu@itp.ac.cn}
 \affiliation{Kavli Institute for Theoretical Physics China (KITPC)  \\
 Institute of Theoretical Physics,
 Chinese Academy of sciences, Beijing 100080, China  }
 %
%\author{ author for second address}
%\affiliation{
\date{$\ $ e-mail: ylwu@itp.ac.cn}
\begin{abstract}
Based on a maximally symmetric minimal unification hypothesis and a
quantum charge-dimension correspondence principle, it is
demonstrated that each family of quarks and leptons belongs to the
Majorana-Weyl spinor representation of 14-dimensions that relate to
quantum spin-isospin-color charges. Families of quarks and leptons
attribute to a spinor structure of extra 6-dimensions that relate to
quantum family charges. Of particular, it is shown that
10-dimensions relating to quantum spin-family charges form a
motional 10-dimensional quantum space-time with a generalized
Lorentz symmetry SO(1,9), and 10-dimensions relating to quantum
isospin-color charges become a motionless 10-dimensional quantum
intrinsic space. Its corresponding 32-component fermions in the
spinor representation possess a maximal gauge symmetry SO(32). As a
consequence, a maximally symmetric minimal unification model SO(32)
containing three families in ten-dimensional quantum space-time is
naturally obtained by choosing a suitable Majorana-Weyl spinor
structure into which quarks and leptons are directly embedded. Both
resulting symmetry and dimensions coincide with the ones of type I
string and heterotic string SO(32) in string theory.
\end{abstract}

\maketitle

\section{Introduction}

The well-known mysterious puzzles in the standard model of
elementary particle physics contain: why parity is non-invariant
in our world? why the observing matter in our world is consist of
three families of quarks and leptons? why the motion of matter in
our living universe is limited in an 4-dimensional space-time.
Revealing those puzzles may concern more fundamental issues, such
as: what is the basic building blocks of nature? what is the basic
symmetries of nature? what is the basic dimensions of space-time?
what is the origin of CP violation and masses? How about the
stability of proton? How about the features of dark matter and
dark energy?

The standard model based on quarks and leptons as basic building
blocks of matter has well been established\cite{SMG,SMW,SMS} and
tested by more and more precise experiments at the energy scale of
order 100 GeV. The electromagnetic interaction is well described by
an U(1) gauge symmetry, and the weak interaction is characterized by
a left-handed $SU(2)_L$ gauge symmetry based on the parity
violation\cite{PV}. The strong interaction of quarks\cite{GM,GZ}
described by the Yang-Mills gauge theory\cite{YM} with a gauge
symmetry $SU(3)_C$ displays a behavior of asymptotic
freedom\cite{AF,AF2}. Three families of quarks and leptons are
necessary for a non-trivial CP-violating phase\cite{KM}, which can
be originated from spontaneous CP violation\cite{CP1,CP2,CP3}. It
has also been proved that the standard model is a renormalizable
quantum field theory\cite{TV}. Of particular, the strength of all
forces has been shown to run into the same magnitude at high energy
scales\cite{RC}, which makes it more attractive for the explorations
of grand unification theory (GUT). The simple and widely
investigated GUT models include $SU(4)\times SU(2)_L\times
SU(2)_R$\cite{PS}, SU(5)\cite{GG} and SO(10)\cite{SO101,SO102}. One
of the important predictions in the GUT models is proton decay. The
current experimental data on proton decays have provided strong
constraints on the minimal SU(5) and SO(10) models.

Unification of families has been much investigated by enlarging
gauge symmetries , such as $O(14)$\cite{O141,O142,O143,O144},
$SO(15)$\cite{SO15}, $SO(16)$\cite{SO16},
$SO(18)$\cite{SO181,SO182,SO183,SO184}, $E_7$\cite{E71,E72} and
$E_8$\cite{E81,E82,E83,E84,E85,E86,E87}. One of the main
difficulties for family unification in four dimensions is the
existence of mirror families at the weak scale\cite{MR}, which was
shown to be avoidable by considering family unification in five and
six dimensional space-time\cite{FU}. As all these models are the
"bottom-up" models within the framework of quantum field theory,
thus the basic building blocks are assumed to be quantum fields of
point particles.

An alternative exploration of unification models is via the
so-called "top-down" approach. String theory\cite{ST1,ST2}, which
was motivated by extending basic building blocks from point
particles to one-dimensional strings, is thought of the most
attractive theory for "top-down" model building. Great efforts have
been made to construct standard like models with three families from
string models\cite{SP}. While it remains an open question at moment
how to uniquely obtain the standard model due to the landscape of
string vacua. Nevertheless, string theory has inspired us with many
interesting concepts. Two of the most important predictions in
string theory are symmetry groups (G) and dimensions (D), i.e.,
G=SO(32) or $E_8 \times E'_8$, and D=10. Especially, the heterotic
string theory\cite{HST1,HST2} has mostly been studied and shown its
potential for a realistic model building.

To arrive at a consistent unification model within a more
fundamental theoretical framework, it is very helpful to explore
unification models via both "bottom-up" and "top-down" approaches.
In this article, we are going to build a general "bottom-up"
unification model that coincides for both symmetry and dimensions
with some "top-down" models in string theory. The paper is organized
as follows: in section 2, we describe in detail the motivations for
introducing a maximally symmetric minimal unification (MSMU)
hypothesis and a quantum charge-dimension correspondence (QCDC)
principle. Then by analyzing the basic quantum charges of quarks and
leptons in the standard model, we show that the minimal basic
building blocks for a general "bottom-up" model should be Majorana
fermions in the spinor representation of (14+6)-dimensions. In
section 3, it is further demonstrated that the spin-family-related
10-dimensions form a motional 10-dimensional quantum space-time and
possess a generalized Lorentz symmetry SO(1,9). Dual to the
spin-family-related 10-dimensions, the isospin-color-related
10-dimensions are found to be motionless and become a 10-dimensional
quantum intrinsic space, its corresponding 32-dimensional spinor
representation is shown to have a maximal gauge symmetry SO(32). We
then arrive at a maximally symmetric minimal unification model
(MSMUM) SO(32) containing three families in ten-dimensional quantum
space-time. It is further shown in section 4 that the family-related
6-dimensions are actually the minimal extra dimensions and will
become a 6-dimensional quantum internal space. Our conclusions and
remarks are presented in the last section with emphasizing the
coincidence for both symmetry and dimensions between the MSMUM
SO(32) and type I or Heterotic String SO(32) in string theory.

\section{Maximally Symmetric Minimal Unification Hypothesis and
Quantum Charge-Dimension Correspondence Principle}

In a general "bottom-up" model building, symmetry should play an
important role. This is because symmetry enables one to establish
relations among different quantum charges of building blocks. It
is well known that basic quantum charges of Dirac fermions are
characterized by spin charges and conjugated charges, which are
described by the Lorentz symmetry SO(1,3). In the standard model,
$SU(2)_L$ symmetry was discovered to describe a symmetry between
two isospin charges of leptons, and $SU(3)_C$ symmetry was
introduced to characterize a symmetry among three color charges of
quarks. $U(1)$ symmetry is known to reflect the electric charge of
particles and antiparticles. The orthogonal symmetry SO(10) was
found to be a unified symmetry for describing isospin-color
charges and their conjugated charges. When treating all the
quantum charges of quarks and leptons on the same
footing\cite{SO14}, the symmetry group $SO(1,3)\times SO(10)$ may
be regarded as a generalized symmetry which describes the
4-dimensional space-time relating to spin-charges and conjugated
charges, and 10-dimensional intrinsic space relating to
isospin-color charges and conjugated charges.

From the above analyzes, it is seen that all symmetries are
introduced based on the basic quantum charges of building blocks
in the standard model. We are then motivated to propose a quantum
charge-dimension correspondence (QCDC) as our working principle,
namely: dimensions of space and time are directly related to basic
quantum charges of building blocks. In this sense, we may mention
such resulting space and time as {\it quantum space and time}.

In general, the independent degrees of freedom of fermionic
building blocks are characterized by the spinor representation of
relevant quantum space and time which relate to basic quantum
charges of building blocks. It is known that each family of quarks
and leptons contains 64 real independent degrees of freedom, which
belong to the Majorana-Weyl spinor representation of 14-dimensions
that relate to the spin-isospin-color charges. It then raises a
question what a symmetry should be introduced to establish
possible relations among independent degrees of freedom of
building blocks. In the existing GUT models, one has considered
symmetries only among basic quantum charges of building blocks
rather than among independent degrees of freedom of building
blocks. Namely, those introduced symmetries are not large enough
to describe possible relations among independent degrees of
freedom of building blocks.

From usefulness and economic considerations, we are motivated to
make a maximally symmetric minimal unification hypothesis
(MSMU-hypothesis) that states that: {\it all independent degrees
of freedom of basic building blocks should be treated equally on
the same footing and related by a possible maximal symmetry in a
minimal unified scheme.}  A direct deduction from such an
MSMU-hypothesis is: {\it fermions as basic building blocks should
be Majorana fermions in the (real) spinor representation of high
dimensions which is determined by the basic quantum charges of
building blocks. The chirality of basic building blocks should
also be well defined to understand parity non-invariance}. Such a
deduction indicates that the possible dimensions should allow a
spinor representation that can well define both Majorana and Weyl
fermions. It is not difficulty to check that when simultaneously
imposing Majorana and Weyl conditions, the allowed dimensions are
restricted to be at $D = 2 + 4n$ $(n=1,2, \cdots )$, i.e., D = 2,
6, 10, 14, $\cdots$.

Inspired from $SO(1,3)\times SO(10)$ GUT model, the minimal
dimension needed for describing each family is D=14, i.e., each
family of fermions belongs to the Majorana spinor representation
of 14-dimensions and has $128=2^{7}$ independent real degrees of
freedom, which are twice to the ones in each family of quarks and
leptons in the standard model. In fact, each family of quarks and
leptons contains 64 real independent degrees of freedom and
belongs to the Majorana-Weyl spinor representation of
14-dimensions.

We now come to the issue on the puzzle of three families of quarks
and leptons. As all three families have the same quantum charges
except their mass hierarchies, it is natural to take three
family-charges and conjugated charges to be basic quantum charges.
Then by applying for the QCDC-principle, the three family-charges
and the conjugated charges must relate to extra 6-dimensions.

As a consequence, the minimal basic building blocks in our general
"bottom-up" unification model should be Majorana fermions in the
spinor representation of (14+6)-dimensions.

\section{Explicit Construction of MSMUM SO(32) in Ten Dimensions}

We shall explicitly demonstrate in this section that when choosing a
suitable spinor structure into which quarks and leptons as building
blocks of standard model are directly embedded, the Majorana
condition in the spinor representation of (14 + 6)-dimensions will
naturally lead to an interesting "bottom-up" MSMUM with a
generalized Lorentz symmetry SO(1,9) in a motional 10-dimensional
quantum space-time and a gauge symmetry SO(32) for each family in
the spinor representation of a motionless 10-dimensional quantum
intrinsic space. Of particular, it will be shown that the motional
10-dimensional quantum space-time is related to the quantum
spin-family charges, and the motionless 10-dimensional quantum
intrinsic space is related to the quantum isospin-color charges.

Let us now construct in detail a general "bottom-up" model based on
the MSMU-hypothesis and QCDC-principle. Denoting $\hat{\Psi}$ to be
the fermionic building block in the spinor representation of
(14+6)-dimensions, the Majorana condition implies that
\begin{equation}
\hat{\Psi} = \hat{\Psi}^{\hat{c}} = \hat{C} \bar{\Psi}^{T}
\end{equation}
where $\hat{\Psi}^{\hat{c}}$ defines the charge conjugation in
(14+6)-dimensions. $\hat{C}$ is the charge conjugate matrix and
satisfies $\hat{C}^{\dagger} = \hat{C}^{-1} = -\hat{C}^{T} =
-\hat{C}$ and $\hat{C}\hat{C}^{\dagger}=1$.

As the first step of a general "bottom-up" model building, it is
crucial to find out an appropriate spinor structure of Majorana
fermions in the spinor representation of high dimensions. This is
because different spinor structures of building blocks reflect
different geometries of space-time, which then result in different
physics phenomena. As a natural and realistic consideration, a
spinor structure of building blocks should be chosen in such a way
that quarks and leptons as building blocks of standard model must
directly be embedded into such a spinor structure. For this
purpose, it turns out that the $8\times 128$-dimensional spinor
representation of Majorana fermion $\hat{\Psi}$ in
(14+6)-dimensions takes the following spinor structure
\begin{equation}
\hat{\Psi} = \left( \begin{array}{c} {\bf \Psi} + i {\bf \tilde{\Psi}} \\
{\bf \Psi} - i {\bf \tilde{\Psi}}
\end{array} \right)
\end{equation}
where ${\bf \Psi}^{T} = (\Psi_1, \Psi_2, \Psi_3, \Psi_0 )^{T}$ and
${\bf \tilde{\Psi}}^{T} = (\tilde{\Psi}_1, \tilde{\Psi}_2,
\tilde{\Psi}_3, \tilde{\Psi}_0 )^{T}$ with $\Psi_i$ and
$\tilde{\Psi}_i$ (i=1,2,3,0) being the fermionic building block in
the spinor representation of 14-dimensions. They satisfy the
Majorana condition in 14-dimensions
\begin{eqnarray}
\Psi_i = \Psi_i^{c} = C \bar{\Psi}_i^{T}, \qquad \tilde{\Psi}_i =
\tilde{\Psi}_i^{c} = C \bar{\tilde{\Psi}}_i^{T}
\end{eqnarray}
Where $\Psi_i^{C}$ ( $\tilde{\Psi}_i^{C}$ ) define the charge
conjugation in 14-dimensions, and the charge conjugate matrix $C$
also satisfies $C^{\dagger}  = C^{-1} = -C^{T} = -C$ and
$CC^{\dagger}=1$. $\Psi_i$ and $\tilde{\Psi}_i$ (i=1,2,3,0) are
Majorana fermions in the 128-dimensional spinor representation of
14-dimensions. The spinor structures ${\bf \Psi} + i {\bf
\tilde{\Psi}}$ and ${\bf \Psi} - i {\bf \tilde{\Psi}}$ in the spinor
representation of extra 6-dimensions are corresponding to the $4$-
and $\bar{4}$-representations of SU(4) which characterizes a family
symmetry for the basic building blocks. Here we have considered the
facts that $SO(6)$ is isomorphic to $SU(4)$ and all families have
the same spin-isospin-color charges. The explicit spinor structure
of Majorana fermions $\Psi_i$ and $\tilde{\Psi}_i$ (i=1,2,3,0) in
14-dimensions is given by the following form
\begin{equation}
\Psi_i = \left( \begin{array}{c} F_{iL} + F'_{iR} \\ F_{iR} +
F'_{iL}
\end{array} \right)
\end{equation}
with $F_{iL,R}$ being defined as
\begin{eqnarray}
F_{iL,R}^{T} & & = [\ U_{r}, U_{b}, U_{g}, N, D_{r}, D_{b}, D_{g},
E,  \nonumber \\
& & D_{r}^{c}, D_{b}^{c}, D_{g}^{c}, E^{c}, -U_{r}^{c}, -U_{b}^{c},
-U_{g}^{c}, -N^{c}\  ]_{i\ L,R}^{T} \nonumber \\
F_{i L,R}^{'T} & & =  [\ U'_{r}, U'_{b}, U'_{g}, N', D'_{r},
D'_{b}, D'_{g}, E',  \\
& & D_{r}^{'c}, D_{b}^{'c}, D_{g}^{'c}, E^{'c}, -U_{r}^{'c},
-U_{b}^{'c}, -U_{g}^{'c}, -N^{'c} \ ]_{i\ L,R}^{T} \nonumber
\end{eqnarray}
where the subindexes 'r', 'b', 'g' denote three colors. Similar
forms hold for $\tilde{\Psi}_i$ with just replacing $F_{iL,R}$ by
$\tilde{F}_{iL,R}$. All the fermions $\psi = U,\ D,\ E,\ N,
\tilde{U},\ \tilde{D},\  \tilde{E},\ \tilde{N},\ \cdots $ are four
complex component Dirac fermions defined in the 4-dimensions. The
indices ``L'' and ``R'' denote the left-handed and right-handed
fermions in 4-dimensions, i.e., $\psi_{L,R} = \frac{1}{2}(1\mp
\gamma_{5}) \psi$. The index ``c'' represents the charge conjugation
in 4-dimensions, $\psi^{c} = c\bar{\psi}^{T}$ with
$c=i\gamma_{0}\gamma_{2}=i\sigma_{3}\otimes \sigma_{2}$.

With the above explicitly given spinor structure of Majorana
fermions in (14+6)-dimensions, it is not difficult to find out the
charge conjugate matrix $\hat{C}$ in (14+6)-dimensions
\begin{eqnarray}
\hat{C} = i \sigma_1\otimes 1 \otimes 1 \otimes \sigma_{1}\otimes
\sigma_{2}\otimes \sigma_{2} \otimes 1\otimes 1 \otimes \sigma_{3}
\otimes \sigma_{2},
\end{eqnarray}
and the corresponding charge conjugate matrix $C$ in 14-dimensions
\begin{eqnarray}
C = i\sigma_{1}\otimes \sigma_{2}\otimes \sigma_{2} \otimes 1\otimes
1 \otimes \sigma_{3} \otimes \sigma_{2}
\end{eqnarray}

From the above spinor structure of Majorana fermions in
(14+6)-dimensions and the charge conjugate matrices, we can
explicitly write down twenty gamma matrices
$\hat{\Gamma}_{\hat{A}}=(\hat{\gamma}_{A}, \Gamma_{I}) =
(\gamma_{a}, \hat{\gamma}_m , \Gamma_{I})$ corresponding to
(4+6+10)-dimensions.  Here $\hat{\gamma}_A = (\gamma_{a},
\hat{\gamma}_m )$ $(a=0,1,2,3; A=0,1,\cdots, 9)$ are the gamma
matrices in 10-dimensions relating to the spin-family charges and
conjugated charges. Their explicit forms are given by
\begin{eqnarray}
\gamma_{0} & = & 1\otimes 1\otimes 1\otimes 1\otimes 1\otimes
1\otimes 1\otimes 1
\otimes \sigma_{1}\otimes 1\ , \nonumber \\
\gamma_{1} & = & i\ 1\otimes 1\otimes 1\otimes 1\otimes 1\otimes
1\otimes 1\otimes 1
\otimes \sigma_{2}\otimes \sigma_{1}\ , \nonumber \\
\gamma_{2} & = & i\ 1\otimes 1\otimes 1\otimes 1\otimes 1\otimes
1\otimes 1\otimes 1
\otimes \sigma_{2}\otimes \sigma_{2}\ , \nonumber \\
\gamma_{3} & = & i\ 1\otimes 1\otimes 1\otimes 1\otimes 1\otimes
1\otimes 1\otimes 1
\otimes \sigma_{2}\otimes \sigma_{3}\ , \nonumber \\
\hat{\gamma}_{4} & = & i\ \sigma_1 \otimes 1\otimes \sigma_2 \otimes
1\otimes 1\otimes 1\otimes 1\otimes 1 \otimes \sigma_{3}\otimes 1 \ , \nonumber \\
\hat{\gamma}_{5} & = & i\ \sigma_2 \otimes \sigma_3 \otimes \sigma_2
\otimes 1\otimes 1\otimes 1\otimes 1\otimes 1 \otimes \sigma_{3}\otimes 1 \ , \nonumber \\
\hat{\gamma}_{6} & = & i\ \sigma_1 \otimes \sigma_2 \otimes \sigma_3
\otimes 1\otimes 1\otimes 1\otimes 1\otimes 1 \otimes \sigma_{3}\otimes 1 \ , \nonumber \\
\hat{\gamma}_{7} & = & i\ \sigma_2 \otimes \sigma_2 \otimes 1
\otimes
1\otimes 1\otimes 1\otimes 1\otimes 1 \otimes \sigma_{3}\otimes 1 \ , \nonumber \\
\hat{\gamma}_{8} & = & i\ \sigma_1 \otimes \sigma_2 \otimes \sigma_1
\otimes 1\otimes 1\otimes 1\otimes 1\otimes 1 \otimes \sigma_{3}\otimes 1 \ , \nonumber \\
\hat{\gamma}_{9} & = & i\ \sigma_2 \otimes \sigma_1 \otimes
\sigma_2 \otimes 1\otimes 1\otimes 1\otimes 1\otimes 1 \otimes
\sigma_{3}\otimes 1
\end{eqnarray}
and $\Gamma_I$ $(I=1,2,\cdots, 10)$ are the gamma matrices in
10-dimensions relating to the isospin-color charges and conjugated
charges. Their explicit forms read
\begin{eqnarray}
\Gamma_{1} & = & \ 1\otimes 1\otimes 1\otimes \sigma_{1}\otimes
\sigma_{1}\otimes 1
\otimes 1\otimes \sigma_{2}\otimes \sigma_{3}\otimes 1\ , \nonumber \\
\Gamma_{2} & = & \ 1\otimes 1\otimes 1\otimes \sigma_{1}\otimes
\sigma_{2}\otimes 1
\otimes \sigma_{3}\otimes \sigma_{2}\otimes \sigma_{3}\otimes 1\ , \nonumber \\
\Gamma_{3} & = & \ 1\otimes 1\otimes 1\otimes \sigma_{1}\otimes
\sigma_{1}\otimes 1 \otimes \sigma_{2}\otimes \sigma_{3}\otimes
\sigma_{3}\otimes 1\ , \nonumber \\
\Gamma_{4} & = & \ 1\otimes
1\otimes 1\otimes \sigma_{1}\otimes \sigma_{2}\otimes 1
\otimes \sigma_{2}\otimes 1\otimes \sigma_{3}\otimes 1\ , \nonumber \\
\Gamma_{5} & = & \ 1\otimes 1\otimes 1\otimes \sigma_{1}\otimes
\sigma_{1}\otimes 1
\otimes \sigma_{2}\otimes \sigma_{1}\otimes \sigma_{3}\otimes 1\ , \nonumber \\
\Gamma_{6} & = & \ 1\otimes 1\otimes 1\otimes \sigma_{1}\otimes
\sigma_{2}\otimes 1
\otimes \sigma_{1}\otimes \sigma_{2} \otimes \sigma_{3}\otimes 1\ , \nonumber \\
\Gamma_{7} & = & \ 1\otimes 1\otimes 1\otimes \sigma_{1}\otimes
\sigma_{3}\otimes \sigma_{1}
\otimes 1\otimes 1\otimes \sigma_{3}\otimes 1\ , \nonumber \\
\Gamma_{8} & = & \ 1\otimes 1\otimes 1\otimes  \sigma_{1}\otimes
\sigma_{3}\otimes \sigma_{2}
\otimes 1\otimes 1 \otimes \sigma_{3}\otimes 1\ , \nonumber \\
\Gamma_{9} & = & \ 1\otimes 1\otimes 1\otimes \sigma_{1}\otimes
\sigma_{3}\otimes \sigma_{3}
\otimes 1\otimes 1\otimes \sigma_{3}\otimes 1\ , \nonumber \\
\Gamma_{10} & = & \ 1\otimes 1\otimes 1\otimes \sigma_{2}\otimes
1\otimes 1\otimes 1\otimes 1\otimes \sigma_{3}\otimes 1\
\end{eqnarray}
where $\sigma_{i}$ ($i=1,2,3$) are Pauli matrices, and ``1'' is
understood as a $2\times 2$ unit matrix. From the above gamma
matrices, one may define three type of chiral gamma matrices
\begin{eqnarray}
\hat{\gamma}_{11} & = &  \hat{\gamma}_{0} \cdots \hat{\gamma}_{9}
= - \sigma_3 \otimes 1\otimes 1\otimes 1 \otimes 1\otimes 1\otimes
1\otimes 1 \otimes
\sigma_3 \otimes 1 \nonumber \\
\Gamma_{15} & = & -i\gamma_{0} \cdots \gamma_3 \Gamma_1 \cdots
\Gamma_{10} = 1\otimes 1\otimes 1\otimes
\sigma_{3}\otimes 1\otimes 1\otimes 1\otimes 1 \otimes \sigma_3 \otimes 1  \nonumber \\
\Gamma_{11} & = & \Gamma_{1} \cdots \Gamma_{10} = 1\otimes
1\otimes 1\otimes \sigma_{3}\otimes 1\otimes 1\otimes 1\otimes 1
\otimes 1\otimes 1
\end{eqnarray}
with $\hat{\gamma}_{11}\hat{\gamma}_{A} = -
\hat{\gamma}_{A}\hat{\gamma}_{11}$, $\Gamma_{15}\Gamma_{I} = -
\Gamma_{I}\Gamma_{15}$, $\Gamma_{15}\hat{\gamma}_{A} = -
\hat{\gamma}_{A}\Gamma_{15}$ and $\Gamma_{11}\Gamma_{I} = -
\Gamma_{I}\Gamma_{11}$.

The Majorana-Weyl representation of $\Psi_i$ in 14-dimensions is
then defined through the projection operators
$P_{W,E}=(1\mp\Gamma_{15})/2$ with $P_{W,E}^{2} = P_{W,E}$
\begin{eqnarray}
\Psi_{Wi} & = & P_{W}\Psi_i = \frac{1}{2}(1 - \Gamma_{15})\Psi_i
\equiv \left( \begin{array}{c} F_{Li} \\ F_{Ri}
\end{array} \right),  \nonumber \\
\Psi_{Ei} & = & P_{E}\Psi_i = \frac{1}{2}(1 + \Gamma_{15})\Psi_i
\equiv
 \left( \begin{array}{c} F'_{Ri}
\\ F'_{Li}
\end{array} \right)
\end{eqnarray}
In order to distinguish with the left- and right-handed fermions
defined via the projection operators $P_{L,R} = (1 \mp
\gamma_{5})/2$ in 4-dimensional space-time, the Majorana-Weyl
fermions $\Psi_{Wi}$ and $\Psi_{Ei}$ are mentioned as 'westward'
and 'eastward' fermions. Similar considerations are applicable to
$\tilde{\Psi}_{i}$. It is of interest to notice the following
fact: each westward Majorana-Weyl fermion $\Psi_{Wi}$ contains 64
independent degrees of freedom, which exactly represent 64
independent degrees of freedom of quarks and leptons in each
family when including the right-handed neutrino. The eastward
Majorana-Weyl fermions $\Psi_{Ei}$ are regarded as mirror
particles, i.e., mirroquarks and mirroleptons.

It is observed that under the transformation of parity and charge
conjugation, the gamma matrices $\hat{\gamma}_{A}$ and $\Gamma_I$
transform with an opposite sign, i.e.,
 \begin{eqnarray}
 \hat{C}  \hat{\gamma}_{A} \hat{C}^{\dagger} = - \hat{\gamma}_{A}^T, \qquad \gamma_0
 \hat{\gamma}_{A} \gamma_0 = \hat{\gamma}_{A}^{\dagger} \nonumber \\
\hat{C} \Gamma_I \hat{C}^{\dagger} = \Gamma_I^T, \qquad \gamma_0
 \Gamma_I \gamma_0 = - \Gamma_I^{\dagger}
 \end{eqnarray}
which indicates that the 10-dimensions relating to the quantum
spin-family charges must be different from the 10-dimensions
relating to the quantum isospin-color charges. Such a difference
should directly reflect their geometries. The former turns out to be
motional with a nontrivial kinetic term and forms a high dimensional
quantum space-time, the latter is found to be motionless and becomes
a quantum intrinsic space. This can be seen more explicitly from the
following identities
\begin{eqnarray}
& & \bar{\hat{\Psi}} \hat{\gamma}^{A}i\partial_{A} \hat{\Psi} \equiv
\frac{1}{2} [ \bar{\hat{\Psi}} \hat{\gamma}^{A}i\partial_{A}
\hat{\Psi} - i\partial_{A} (\bar{\hat{\Psi}}) \hat{\gamma}^{A}
 \hat{\Psi} ] \nonumber \\
& & \bar{\hat{\Psi}} \Gamma^{I}i\partial_{I} \hat{\Psi} \equiv
\frac{1}{2} \partial_{I} \left( \bar{\hat{\Psi}} i \Gamma^{I}
\hat{\Psi} \right)
\end{eqnarray}
with $A=0,1,\dots, 9$, and $I = 1, \cdots , 10$. In obtaining the
above identities, we have used the Majorana condition $\hat{\Psi}=
\hat{\Psi}^{\hat{c}}$. As the second identity is given by a total
derivative, which means that no kinetic term can exist in
10-dimensions relating to the isospin-color charges. Namely, the
corresponding space is motionless and will be mentioned as an
10-dimensional quantum intrinsic space.

As the spin-family-related 10-dimensional quantum space-time is
motional, it possesses a generalized maximal Lorentz symmetry
SO(1,9). In contrast, since the isospin-color-related 10-dimensional
quantum intrinsic space is motionless, its 32-component fermions in
the spinor representation turn out to get a maximal symmetry
$SO(32)$ when they are moving in 10-dimensional quantum space-time.
Note that in a motional 4-dimensional space-time without considering
three families and introducing extra 6-dimensions relating to
three-family charges, the maximal gauge symmetry was found to be a
unitary symmetry SU(32)\cite{SU32}.

The generators of the symmetry group SO(32) are given by the
following tensors constructed from $\Gamma$-matrices
\begin{eqnarray}
& & T^{U} \equiv (\Gamma_{11},\  \Sigma^{IJ}, \  \Gamma_{11}
\Sigma^{IJKL} ),\qquad T_5^{V} \equiv (\Sigma^{IJK},\ i\Gamma_{11}
\Sigma^{IJK}) \nonumber \\
& & T^{\hat{U}} \equiv (T^{U}, \ i\Gamma_{15} T_5^{V}), \qquad
(\hat{U}=1,\cdots, 496 )
\end{eqnarray}
where $\Sigma^{IJ} = \frac{i}{4} [ \Gamma^I , \Gamma^J ]$ is a
two-rank antisymmetric tensor which forms the generators of subgroup
SO(10), and others are the high-rank antisymmetric tensors with
$T^U$ ($U=1, \cdots, 256$) being the generators of subgroup U(16).
The generators of the generalized Lorentz symmetry SO(1,9) in
10-dimensional quantum space-time are given by a two-rank
antisymmetric tensor of gamma matrices
\begin{eqnarray}
\Sigma^{AB} = \frac{1}{4i} [ \hat{\gamma}^A , \hat{\gamma}^B ]
\end{eqnarray}
Under the charge conjugation and parity transformation, the tensors
of gamma matrices transform as follows
\begin{eqnarray}
& & \hat{C} \Sigma^{AB} \hat{C}^{\dagger} = C \Sigma^{AB}
C^{\dagger}= - (\Sigma^{AB}) ^{T}, \quad  \gamma^0
\Sigma^{AB}\gamma^0 = (\Sigma^{AB})^{\dagger}, \nonumber   \\
 & & \hat{C} T^{\hat{U}} \hat{C}^{\dagger} = C T^{\hat{U}} C^{\dagger}= - (T^{\hat{U}})
^{T}, \quad  \gamma^0 T^{\hat{U}} \gamma^0 = (T^{\hat{U}})^{\dagger}
\end{eqnarray}

We are now in the position to explicitly construct an Lagrangian
for the general "bottom-up" MSMUM with a gauge symmetry SO(32) and
a generalized Lorentz symmetry SO(1,9) in the motional
10-dimensional space-time. For our purpose in this article, the
10-dimensional space-time is treated as a flat space-time without
considering gravity. With the Majorana-Weyl fermions in the
10-dimensional space-time, the Lagrangian is found to be
\begin{eqnarray}
{\cal L}^{10D} & = & \frac{1}{4}\bar{\hat{\Psi}} \hat{\gamma}^{A}i
D_{A} \hat{\Psi}- \frac{1}{4 g_U^2} {\cal F}_{AB}^{\hat{U}} {\cal
F}^{\hat{U} AB}
\end{eqnarray}
which is self-Hermitian due to Majorana condition. $D_{A}$ is the
covariant derivative corresponding to the gauge field ${\cal
A}_{A}^{\hat{U}}$ ($A=0,\cdots, 9$) in 10-dimensions
\begin{eqnarray}
D_{A} = \partial_{A} - i g_U {\cal A}_{A}^{\hat{U}} T^{\hat{U}}
\end{eqnarray}
and its commutation relation defines the field strength ${\cal
F}_{AB}^{\hat{U}}$
\begin{eqnarray}
i[ D_A , D_B ] = {\cal F}_{AB}^{\hat{U}}  T^{\hat{U}}
\end{eqnarray}
with $T^{\hat{U}}$ $(\hat{U}=1,\cdots, 496)$ being the generators of
SO(32) and $g_U$ the coupling constant.

To maintain four-dimensional Poincare invariance and fit the
so-called no-go theorem\cite{CM}, we shall take the 10-dimensional
space-time to be of the form $M^4 \times K$, where $M^4$ is
4-dimensional Minkowski space and $K$ is a compact 6-dimensional
internal space. The symmetry in the product space $M^4 \times K$ is
corresponding to $SO(1,3) \times SO(6)$ which is a subgroup of
SO(1,9).

In the 10-dimensional quantum space-time, one can define the
Majorana-Weyl fermions via the projection operators $P_{\pm}=(1\pm
\hat{\gamma}_{11})/2$ with $P_{\pm}^{2} = P_{\pm}$
\begin{eqnarray}
\hat{\Psi}_{\pm} = P_{\pm}\hat{\Psi} = \frac{1}{2}(1 \pm
\hat{\gamma}_{11} ) \hat{\Psi}
= \left( \begin{array}{c} {\bf \Psi}_{L,R} + i {\bf \tilde{\Psi}}_{L,R} \\
{\bf \Psi}_{R,L} - i {\bf \tilde{\Psi}}_{R,L}
\end{array} \right)
\end{eqnarray}
where
\begin{eqnarray}
& & {\bf \Psi}_{L}=\left( \begin{array}{c} \textbf{F}_{L}
\\ \textbf{F}'_{L}
\end{array} \right), \qquad  {\bf \Psi}_{R}=\left( \begin{array}{c}
 \textbf{F}'_{R} \\ \textbf{F}_{R}
\end{array} \right)
\end{eqnarray}
Similar forms hold for ${\bf \tilde{\Psi}}_{L,R}$ with replacing
$\textbf{F}_{L,R}$ and $\textbf{F}'_{L,R}$ by
$\tilde{\textbf{F}}_{L,R}$ and $\tilde{\textbf{F}}'_{L,R}$
respectively. Where $\textbf{F}^T_{L,R} = (F_1,F_2,F_3,F_0)_{L,R}$
and $\tilde{\textbf{F}}^T_{L,R} = (\tilde{F}_1, \tilde{F}_2,
\tilde{F}_3, \tilde{F}_0)_{L,R}$ are two kinds of fermionic building
blocks, the corresponding $\textbf{F}'_{L,R}$ and
$\tilde{\textbf{F}}'_{L,R}$ are their mirror fermionic building
blocks. Here $F_{iL,R}$ ($F'_{iL,R}$) and $\tilde{F}_{iL,R}$
($\tilde{F}'_{iL,R}$) (i=1,2,3,0) belong to the spinor
representation in (4+8)-dimensions.

It is noticed that the 10-dimensional Weyl fermions $\hat{\Psi}_{-}$
and $\hat{\Psi}_{+}$ can be related by a kind of F-parity operation
$\hat{F}$ defined as follows
\begin{eqnarray}
& &  \hat{\Psi}_{-}= \hat{\Psi}_{+}^{\hat{F}} = \hat{\Gamma}_0
\hat{\Psi}_{+} (\tilde{\Psi} \to -
\tilde{\Psi}) \\
& & \hat{\Gamma}_0 = \sigma_1 \otimes 1\otimes 1\otimes 1 \otimes
1\otimes 1\otimes 1\otimes 1 \otimes 1 \otimes 1 \nonumber
\end{eqnarray}
which implies that one may take one of the 10-dimensional
Majorana-Weyl fermions $\hat{\Psi}_{+}$ as the basic building blocks
for constructing the Lagrangian, namely
\begin{eqnarray}
{\cal L}^{10D} & = & \frac{1}{2}\bar{\hat{\Psi}}_{+}
\hat{\gamma}^{A}i D_{A} \hat{\Psi}_{+}- \frac{1}{4 g_U^2} {\cal
F}_{AB}^{\hat{U}} {\cal F}^{\hat{U} AB}
\end{eqnarray}

\section{Minimal Extra 6-dimensions and Quantum Family-charges }

In the above considerations, we have introduced extra 6-dimensions
by relating to three family charges and conjugated charges based on
the QCDC-principle. Here we shall directly demonstrate that the
extra 6-dimensions are actually the minimal ones in addition to the
14-dimensions relating to the spin-isospin-color charges. The reason
can be found from the fact that since the fermionic basic building
blocks in the 128-dimensional spinor representations of
14-dimensions are Majorana fermions, for extra dimensions to be
nontrivial with kinetic term, it requires that the corresponding
gamma matrices in the spinor representation of extra dimensions must
be antisymmmetric. The question then becomes that starting from
which dimension the possible antisymmetric gamma matrices that
satisfy the anticommutation relations can be equal to the
corresponding dimensions. It is not difficult to check that in two
dimensions there is only one antisymmetric gamma matrix (i.e.,
$\sigma_2$), and in four dimensions there are at most three
antisymmetric gamma matrices that satisfy anticommutation relations
(i.e., $\sigma_2 \times \sigma_1$, $\sigma_2 \times \sigma_3$, $1
\times \sigma_2$) . Namely, in two and four dimensions the number of
antisymmetric gamma matrices is less than the number of dimensions.
Only up to six dimensions, there exist six antisymmetric gamma
matrices which satisfy the anticommutation relations. The explicit
forms of six antisymmetric gamma matrices corresponding to six
dimensions can be expressed as follows
\begin{eqnarray}
 &  &\hat{\gamma}_1 = i\ \sigma_1 \otimes 1\otimes \sigma_2  \ , \nonumber \\
 & & \hat{\gamma}_2 = i\ \sigma_3 \otimes \sigma_3 \otimes \sigma_2 \ , \nonumber \\
 &  & \hat{\gamma}_3 = i\ \sigma_1 \otimes \sigma_2 \otimes \sigma_3 \ , \nonumber \\
& & \hat{\gamma}_4 = i\ \sigma_3 \otimes \sigma_2 \otimes 1  \ , \nonumber \\
& & \hat{\gamma}_5 = i\ \sigma_1 \otimes \sigma_2 \otimes \sigma_1\ , \nonumber \\
& & \hat{\gamma}_6 = i\ \sigma_3 \otimes \sigma_1\otimes \sigma_2
%& & \hat{\gamma}_7 = i\ \sigma_2 \otimes 1 \otimes 1 =
%\hat{\gamma}_1 \cdots \hat{\gamma}_6
\end{eqnarray}
which shows that the minimal extra dimensions are truly six
dimensions. As the Lorentz symmetry group SO(6) of extra
6-dimensions is isomorphic to SU(4), the corresponding gamma
matrices can be chosen in a complex spinor representation of
6-dimensions. Thus the gamma matrices corresponding to
spin-family-related 10-dimensions can also be given in a complex
spinor representation, which has actually been constructed
explicitly in the previous section.

With the above explicit demonstrations, we are led to the
conclusion that three family-charges are in fact the minimal basic
quantum charges in addition to the spin-isospin-color charges.
Consequently, the spin-family-related 10-dimensional quantum
space-time becomes a minimal quantum space-time in a general
"bottom-up" MSMUM.

\section{Conclusions and Remarks }

Based on the maximally symmetric minimal unification hypothesis
and the quantum charge-dimension correspondence principle, we have
built, by analyzing the basic quantum charges of quarks and
leptons in the standard model, a maximally symmetric minimal
unification model SO(32) containing three families in ten
dimensional quantum space-time. It is of interest to notice that
both resulting symmetry and dimensions coincide with the ones of
type I and Heterotic string SO(32) in string theory. In addition,
the present motional 10-dimensional quantum space-time gets a
physical meaning as it directly relates to the basic quantum
spin-family charges of quarks and leptons. Also the symmetry group
SO(32) characterizes a maximal gauge symmetry among the
independent degrees of freedom for each family of fermionic
building blocks which belong to the Majorana spinor representation
of 10-dimensional quantum intrinsic space. The 10-dimensional
quantum intrinsic space has turned out to be motionless and
directly relate to the basic quantum isospin-color charges of
quarks and leptons. The quantum spin-family charges and quantum
isospin-color charges appear to be a kind of dual quantum charge.
In the real world, the spin-color charges are conserved by
symmetries, while the isospin-family charges are no longer
conserved as the corresponding symmetries have to be broken down.

It has been seen that the most crucial point in building a general
"bottom-up" MSMUM is to find out a suitable spinor structure into
which quarks and leptons as building blocks of standard model are
directly embedded. It is also interesting to observe that all
interactions in the MSMUM SO(32) are self-Hermitian due to
Majorana condition, thus CP symmetry is preserved in the model. In
general, parity is going to be broken down when quarks and leptons
get different masses with mirroquarks and mirroleptons, and CP
symmetry has to be broken down spontaneously.

Similar to GUT models, the key issue becomes how to obtain the
standard model with three families of quarks and leptons from MSMUM
SO(32) in ten dimensions, which will depend on the compactification
of extra dimensions and the symmetry breaking pattern of SO(32). As
the families are manifestly related to the extra 6-dimensions in the
10-dimensional MSMUM SO(32) with a maximal family symmetry SO(6)
which is isomorphic to SU(4), thus the observing three families of
quarks and leptons must be resulted from SU(4) symmetry broken down
to SU(3) family symmetry, which can be very similar to the
Parti-Salam SU(4) symmetry broken down to SU(3) color symmetry of
strong interaction. More specifically, three families of quarks and
leptons characterized by the spinor fermions $F_{iL}$ (i=1,2,3)
within the spinor structure defined in eqs.(2-5) can simply be
realized as zero modes of extra 6-dimensions. In general, it is more
interest to realize compactifications on 6-dimensional Calabi-Yau
manifolds of SU(3) holonomy.

Before ending, we would like to address that when applying the
MSMU-hypothesis to space-time symmetry, it should naturally lead
to a supersymmetric MSMUM since the supersymmetry is the maximal
symmetry for space-time. It is then intriguing to extend the MSMUM
SO(32) in 10-dimensions to a supersymmetric 10-dimensional MSMUM
SO(32) and study its possible relation with the type I string or
heterotic string SO(32) in string theory.

We have laid in this article only the foundation for obtaining a
natural ''bottom-up" MSMUM SO(32) in ten dimensions. Many puzzles
such as stability of proton, dark matter of universe, origin of
masses and mixing, flavor physics at low energies, all require us
to figure out a suitable compactification of extra 6-dimensions
and symmetry breaking pattern of the gauge group SO(32). We shall
investigate those puzzles and more interesting features of MSMUM
SO(32) elsewhere.

\acknowledgments

\label{ACK}

This work was based on the invited plenary talk given at the
International Conference on String Theory (Strings2006) and
supported in part by the National Science Foundation of China
(NSFC) under the grant 10475105, 10491306, and the Project of
Knowledge Innovation Program (PKIP) of Chinese Academy of
Sciences. He would like to express his thanks to many colleagues
for valuable discussions and conversations during the strings2006
conference.

\end{document}